# Generalized Kirchhoff-Law-Johnson-Noise (KLJN) secure key exchange system using arbitrary resistors

Gergely Vadai*, Robert Mingesz and Zoltan Gingl

Department of Technical Informatics, University of Szeged, Hungary

* e-mail: vadaig@inf.u-szeged.hu

## Abstract

The Kirchhoff-Law-Johnson-Noise (KLJN) secure key exchange system has been introduced as a simple, very low cost and efficient classical physical alternative to quantum key distribution systems. The ideal system uses only a few electronic components – identical resistor pairs, switches and interconnecting wires – to guarantee perfectly protected data transmission. We show that a generalized KLJN system can provide unconditional security even if it is used with significantly less limitations. The more universal conditions ease practical realizations considerably and support more robust protection against attacks. Our theoretical results are confirmed by numerical simulations.

## Introduction

Incredibly huge amount of information is transferred in every second – electronic communication is getting one of the most important parts of the industry, economy, medicine, education, entertainment, personal life and many more. It is obvious that in several cases reliable protection of the information is essential and it is not easy at all. There are many methods and algorithms to encrypt data and although they appear practically unbreakable today, one can't be sure about their reliability in the future.

Natural physical processes and systems can be promising alternatives to digital encryption to hide information from the eavesdroppers. The most is expected from the quantum key distribution (QKD); however an extremely simple and low cost classical physical system has been proposed as an efficient alternative to provide unconditional security using only a few resistors, switches and wires[1]. The idea is surprisingly simple and elegant: the thermal (Johnson) noise of the resistor serves as the signal source and only a simple circuit is needed to establish communication between the two parties, usually named as Alice and Bob. The schematic of the circuit is shown on Figure 1.

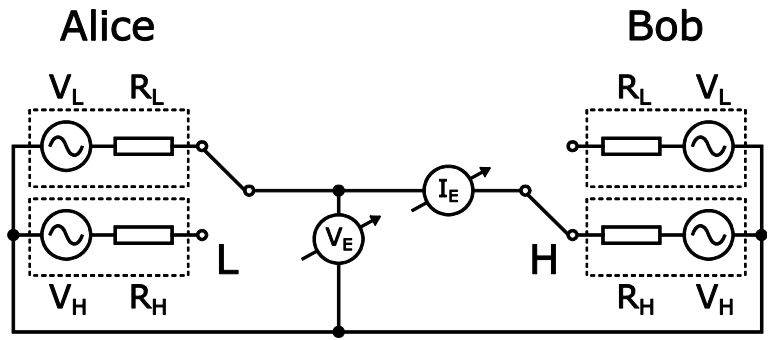

**Figure 1** The KLJN key exchanger uses identical pair of resistors at both ends. The voltage generators represent the thermal noise of the resistors. LH state is shown.

At both ends of the communication line there is an identical pair of resistors with publicly known lower and higher values, $R_L$ and $R_H$, respectively. Both communicating parties can connect one of the resistors to the interconnecting wire. It has been shown using the second law of thermodynamics[1] and mathematical statistics[2] that the eavesdropper can't distinguish between the two cases when different resistors are selected at the two ends, therefore one bit of information can be transferred with unconditional security. Note that the zero-mean Johnson noise voltages of the resistors are represented by voltage generators. The power spectral density of the thermal noise voltage is equal to $4kTR$, where k is the Boltzmann constant and $T$ is the temperature of the resistor $R$. The use of separate voltage generators allows high enough voltage amplitudes in practical applications, virtually emulates very high temperatures. Recently it has been proven using mathematical statistical methods that only this kind of noise can guarantee security[3].

The outstanding performance of the KLJN system has been demonstrated by the first practical realization using a digital signal processor based system[4] and has inspired the development of another secret key exchange method[5]. The simplicity and flexibility of the system allowed the proposal of many possible applications in different environments including securing computer communications, hardware components, memories, processors, keyboards, mass storage devices, key distribution over the Smart Grid, ethernet cables, uncloneable hardware keys and automotive communication[6-9].

The ideal KLJN key exchange system is considered to be unconditionally secure; however numerous attacks were introduced using the non-ideal properties that appear in practical realizations. The tolerance of resistor values and temperature, the resistance of the interconnecting wires and switches, any deviation from the ideal noise distribution and even interference can cause information leak[10-20]. The latest attack is based on directional wave measurements and uses the finite resistance of the transmission wire[21,23]. There is a continuous debate about the validity and limitations about these attack methods.

Any practical realization has limitations and exhibits non-ideal behaviour, therefore the question arises, how sensitive the security on the tolerance of the properties of the components is; how much the ideal cases can be approximated; are there any efficient methods to compensate the effects of non-ideal features?

Here we generalize the KLJN key exchange protocol by using arbitrary resistors and we prove that in the ideal case it is still unconditionally secure, if the noise voltages are properly chosen. This allows operating the practical realizations close to the ideal one and in the same time makes flexible, even real-time compensation techniques possible. We have carried out numerical simulations to confirm our theoretical results.

# Results

## The generalized KLJN key exchange protocol

In the original KLJN system two identical resistor pairs are used at both ends of the communication wire. Here we allow the use of four arbitrarily chosen resistors as shown on Figure 2.

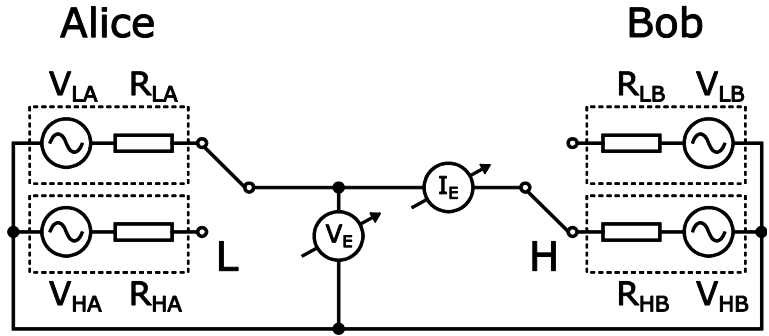

**Figure 2** The generalized KLJN key exchanger uses four different resistors and noise voltage generators. LH state is shown.

The Johnson noise of the resistors is represented by separate voltage generators again. The eavesdropper Eve can measure the current $I_E(t)$ and the voltage $V_E(t)$ in the interconnecting wire and can process these signals to determine which resistors are connected to the wire. If in both the LH and HL cases the same information is obtained, then no information leak occurs. Table 1 summarizes the possibilities.

| State | Alice | Bob | $I_E(t)$ | $V_E(t)$ |
|---|---|---|---|---|
| LH | $R_{LA}, V_{LA}$ | $R_{HB}, V_{HB}$ | $\dfrac{V_{HB}(t) - V_{LA}(t)}{R_{LA} + R_{HB}}$ | $\dfrac{R_{HB}V_{LA}(t) + R_{LA}V_{HB}(t)}{R_{LA} + R_{HB}}$ |
| HL | $R_{HA}, V_{HA}$ | $R_{LB}, V_{LB}$ | $\dfrac{V_{LB}(t) - V_{HA}(t)}{R_{HA} + R_{LB}}$ | $\dfrac{R_{LB}V_{HA}(t) + R_{HA}V_{LB}(t)}{R_{HA} + R_{LB}}$ |

**Table 1** Selected components and quantities in LH and HL states.

## Conditions for security

The eavesdropper can measure the variance of the voltage and current and the correlation of these signals. All these must be the same for both the LH and HL cases.

First, consider the variance of the current in the LH and HL states:

$$\left\langle I_{E,LH}^2(t) \right\rangle = \left\langle I_{E,HL}^2(t) \right\rangle. \tag{1}$$

Substituting the expressions of $I_E$ given in Table 1 we get:

$$\frac{\langle V_{LA}^2(t)\rangle + \langle V_{HB}^2(t)\rangle}{(R_{LA}+R_{HB})^2} = \frac{\langle V_{HA}^2(t)\rangle + \langle V_{LB}^2(t)\rangle}{(R_{HA}+R_{LB})^2}. \quad (2)$$

The equivalence of the variance of the voltage fluctuations observed by Eve implies that

$$\langle V_{E,LH}^2(t)\rangle = \langle V_{E,HL}^2(t)\rangle. \quad (3)$$

Again, expanding these using the formulas for $V_E$ provided in Table 1 yields

$$\frac{R_{HB}^2\langle V_{LA}^2(t)\rangle + R_{LA}^2\langle V_{HB}^2(t)\rangle}{(R_{LA}+R_{HB})^2} = \frac{R_{LB}^2\langle V_{HA}^2(t)\rangle + R_{HA}^2\langle V_{LB}^2(t)\rangle}{(R_{HA}+R_{LB})^2}. \quad (4)$$

Finally, it is required to have the same correlation between the current and voltage signals:

$$\langle I_{E,LH}(t)V_{E,LH}(t)\rangle = \langle I_{E,HL}(t)V_{E,HL}(t)\rangle. \quad (5)$$

Substituting the current and voltage expressions from Table 1 into Eqn. 5 results the following equation:

$$\left\langle \frac{V_{HB}(t)-V_{LA}(t)}{R_{LA}+R_{HB}} \cdot \frac{R_{HB}V_{LA}(t)+R_{LA}V_{HB}(t)}{R_{LA}+R_{HB}} \right\rangle = \left\langle \frac{V_{LB}(t)-V_{HA}(t)}{R_{HA}+R_{LB}} \cdot \frac{R_{LB}V_{HA}(t)+R_{HA}V_{LB}(t)}{R_{HA}+R_{LB}} \right\rangle. \quad (6)$$

The left hand side of this formula can be written as

$$\left\langle \frac{R_{LA}V_{HB}^2(t)+R_{HB}V_{HB}(t)V_{LA}(t)-R_{LA}V_{HB}(t)V_{LA}(t)-R_{HB}V_{LA}^2(t)}{(R_{HA}+R_{LB})^2} \right\rangle =$$
$$= \frac{R_{LA}}{(R_{LA}+R_{HB})^2}\langle V_{HB}^2(t)\rangle + \frac{R_{HB}-R_{LA}}{(R_{LA}+R_{HB})^2}\langle V_{HA}(t)V_{LB}(t)\rangle - \frac{R_{HB}}{(R_{LA}+R_{HB})^2}\langle V_{LA}^2(t)\rangle \quad (7)$$
$$= \frac{R_{LA}}{(R_{LA}+R_{HB})^2}\langle V_{HB}^2(t)\rangle - \frac{R_{HB}}{(R_{LA}+R_{HB})^2}\langle V_{LA}^2(t)\rangle$$

and for the right hand side of Eqn. 6 we obtain

$$\left\langle \frac{R_{HA}V_{LB}^2(t)+R_{LB}V_{HA}(t)V_{LB}(t)-R_{HA}V_{HA}(t)V_{LB}(t)-R_{LB}V_{HA}^2(t)}{(R_{HA}+R_{LB})^2} \right\rangle =$$
$$= \frac{R_{HA}}{(R_{HA}+R_{LB})^2}\langle V_{LB}^2(t)\rangle + \frac{R_{LB}-R_{HA}}{(R_{HA}+R_{LB})^2}\langle V_{HA}(t)V_{LB}(t)\rangle - \frac{R_{LB}}{(R_{HA}+R_{LB})^2}\langle V_{HA}^2(t)\rangle \quad (8)$$
$$= \frac{R_{HA}}{(R_{HA}+R_{LB})^2}\langle V_{LB}^2(t)\rangle - \frac{R_{LB}}{(R_{HA}+R_{LB})^2}\langle V_{HA}^2(t)\rangle.$$

We have used the fact that the cross correlation terms are zero, since all voltage noise signals are independent. Using Eqn. 6, 7 and 8 finally we get

$$\frac{R_{LA}}{(R_{LA}+R_{HB})^2}\left\langle V_{HB}^2(t)\right\rangle - \frac{R_{HB}}{(R_{LA}+R_{HB})^2}\left\langle V_{LA}^2(t)\right\rangle = \frac{R_{HA}}{(R_{HA}+R_{LB})^2}\left\langle V_{LB}^2(t)\right\rangle - \frac{R_{LB}}{(R_{HA}+R_{LB})^2}\left\langle V_{HA}^2(t)\right\rangle. \quad (9)$$

This equation is in agreement with the physical argument that the power flow from the two sides must be the same for HL and LH states[20].

In summary, Eqn. 2,4 and 9 must be satisfied in order to guarantee secure communication.

Note here that in the case of the original KLJN protocol both sides of Eqn. 9 are zero and one can easily see that in agreement with the previous results[1-3] the noise variance scales with the value of the corresponding resistor. In our generalized case the voltage and current signals observed by the eavesdropper are not independent, but they are the same for both the LH and HL states.

## Determining the voltage amplitudes

According to the above it is possible to choose four arbitrary resistor values and one of the voltage variances. In the following example we assume that the variance of $V_{LA}(t)$ is given and the rest three must be calculated using Eqn. 2,4 and 9:

$$\left\langle V_{HB}^2(t)\right\rangle = \left\langle V_{LA}^2(t)\right\rangle \frac{R_{LB}(R_{HA}+R_{HB})-R_{HA}R_{HB}-R_{HB}^2}{R_{LA}^2+R_{LB}(R_{LA}-R_{HA})-R_{HA}R_{LA}}, \quad (10)$$

$$\left\langle V_{HA}^2(t)\right\rangle = \left\langle V_{LA}^2(t)\right\rangle \frac{R_{HA}^2+R_{LB}(R_{HB}+R_{HA})+R_{HA}R_{HB}}{R_{LA}^2+R_{LB}(R_{LA}+R_{HB})+R_{HB}R_{LA}}, \quad (11)$$

$$\left\langle V_{LB}^2(t)\right\rangle = \left\langle V_{LA}^2(t)\right\rangle \frac{R_{LB}^2+R_{LB}(R_{HA}-R_{HB})-R_{HA}R_{HB}}{R_{LA}^2+R_{LA}(R_{HB}-R_{HA})-R_{HA}R_{HB}}. \quad (12)$$

It is important to note that since the variance is proportional to the temperature, the restrictions on the noise variances can be interpreted physically as having different temperatures for each resistor.

## Numerical simulations

In order to confirm our theoretical results we have carried out numerical simulations. The variances and the correlation of the current $I_E$ and voltage $V_E$ were calculated for both the HL and LH cases. We have tested three different configurations as listed in Table 2. Three bit error rate (BER) values have been calculated using the variance of the current, variance of the voltage and the correlation data for the transfer of 1 million bits. The BER values are very close to 50% indicating that the leak is practically zero.

| HL | | LH | | $\left\langle I_E^2(t)\right\rangle$ | $\left\langle V_E^2(t)\right\rangle$ | $\left\langle V_E(t)I_E(t)\right\rangle$ |
|---|---|---|---|---|---|---|
| $R_{HA}$ [kOhm] | $R_{LB}$ [kOhm] | $R_{LA}$ [kOhm] | $R_{HB}$ [kOhm] | BER [%] | BER [%] | BER [%] |
| 9 | 1 | 1 | 9 | 49.967 | 49.973 | 49.994 |
| 10 | 5 | 1 | 9 | 49.985 | 49.980 | 49.979 |
| 5 | 5 | 1 | 9 | 49.981 | 49.984 | 49.981 |

**Table 2** Bit error rate for three different configurations calculated using the variance of the current, variance of the voltage and the correlation coefficient. The first data row corresponds to the original KLJN configuration.

Figure 3 and 4 visualizes the statistics of the quantities that can be observed by the eavesdropper in the HL and LH states. It can be easily seen that in both states the histograms are practically identical, the eavesdropper can't extract any information from the data using any of the three considered quantities.

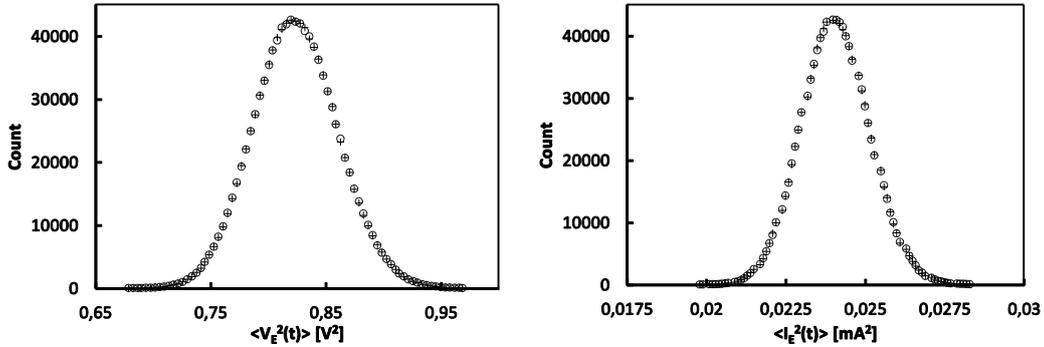

**Figure 3** Histograms of the variance of the voltage $V_E$ and current $I_E$ obtained during the transfer of $10^6$ bits. The HL state ($R_{HA}$=10 kOhm, $R_{LB}$=5 kOhm, $V_{HA}$=2.179 V, $V_{LB}$=0.816 V) is indicated with crosses and the LH state ($R_{LA}$=1 kOhm, $R_{HB}$=9 kOhm, $V_{LA}$=1 V, $V_{HB}$=1.186 V) is marked with circles.

Figure 4 shows the correlation histogram of the current and voltage. The correlation is not zero in contrast to the original KLJN configuration[2,3]. On the right hand side the scatter plot corresponding to the joint probability distribution of the voltage and current can be seen. The asymmetrical shape indicates the non-zero correlation as well[3] and it corresponds to a non-zero, but identical power flow from one side to the other both in the HL and LH cases[20].

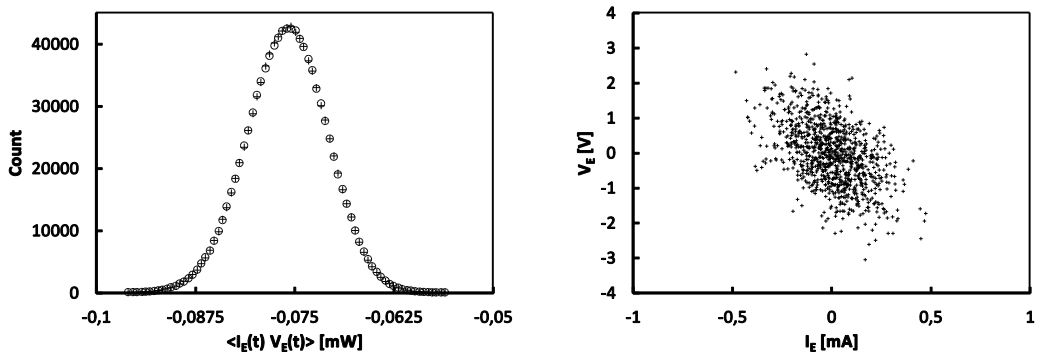

**Figure 4** On the left panel the histograms of the correlation of the voltage $V_E$ and current $I_E$ obtained during the transfer of $10^6$ bits are plotted. The HL state ($R_{HA}$=10 kOhm, $R_{LB}$=5 kOhm, $V_{HA}$=2.179 V, $V_{LB}$=0.816 V) is indicated with crosses and the LH state ($R_{LA}$=1 kOhm, $R_{HB}$=9 kOhm, $V_{LA}$=1 V, $V_{HB}$=1.186 V) is marked with circles. On the right panel the 1000 points of the $V_E$, $I_E$ scatter plot can be seen for a single bit transfer[24].

# Discussion

Our theoretical and numerical simulation results show that the KLJN key exchange protocol provides security even under significantly generalized conditions: the resistor values can be chosen arbitrarily. In order to ensure security the voltage noise variances must be chosen according to the equations we have presented above. This means that the required dependence of voltage amplitudes on resistor values has also been generalized. Considering the original physical picture this means that the system is not any more in thermal equilibrium, the temperature of the resistors will be different. However, since practical realizations use separate artificial noise generators to emulate the Johnson noise of the resistors, this does not mean any limitation or problem. On the other hand Eqn. 9 is equivalent to the assumption that the mean power flow in the two states are the same[20]. The original protocol is a special case in our approach, when both sides of Eqn. 9 equal zero.

The generalized protocol helps to improve the security in practical applications significantly and allows new realizations as well. For example consider the original KLJN system using resistors with a given tolerance. In this case the original protocol will cause some information leak, however if the resistor values are measured, the voltage amplitudes can be slightly changed to eliminate the leak. Note that tuning the voltage amplitude is rather easy to implement by today's precision analogue and digital circuits. It is easy to see, that even the parasitic resistance of switches and the interconnecting wire resistance can be compensated, it is a special case in our approach. This is in agreement with the complete elimination of the effect of wire resistance that has been discussed by Kish et al[20].

During the communication both Alice and Bob can measure the components' resistance continuously; they can measure voltage at different points and can also determine the loop current. The on-line measurement of the communicating resistors, resistance of the switches and resistance of the interconnecting wire is possible, and the parties can inform each other about their observations even via a public link. Therefore it is also possible to implement real-time continuous information leak elimination. For example automatic compensation of the changes in the wire resistance is possible, when for any reason the interconnecting cable is replaced. These features are very important in practical applications, because the sensitivity of the original KLJN system on the non-ideal properties of the components can be dramatically reduced and several attack types can be eliminated.

# Methods

Our calculations are mainly based on the mathematical statistical approach described in our recent publications in the subject[2,3]. We have calculated the variances of the quantities and the correlation between the considered signals. Agreement with previous results based on classical physical approach was also important and it has been demonstrated.

Numerical simulations were based on a code written in the LabVIEW environment and the source code is available for download[24]. The simulation of the transfer of a single bit in the HL and LH state was carried out by generating 1000 samples to represent the four voltage noise signals. Then $V_E$ and $I_E$ observed by the eavesdropper were calculated. Using these data the variances of $V_E$ and $I_E$ and the correlation values were determined.

In order to check any possible information leak and to measure the statistics of the above mentioned quantities 1 million bits were transferred. The BER was calculated using all three indicators. During the BER estimation all bit transfers were taken for both the HL and LH states and the median was used as a threshold value to determine the bit state for an individual bit transfer. After this we have calculated the ratio of unsuccessful bit guessing to get the BER value.

# Author contributions

G.V., R.M. and Z.G. designed the study; G.V. found the theoretical derivation and carried out the numerical simulations; G.V., R.M. and Z.G. discussed and interpreted the results; Z.G. wrote the paper; G.V., R.M. and Z.G. proofed the paper.